\documentclass[prl,twocolumn,showpacs,showkeys,superscriptaddress, nofootinbib]{revtex4-2}
\usepackage{graphicx,amsmath,hyperref,url}
\usepackage{xcolor}
\usepackage{braket}
\usepackage{amsmath, amssymb}
\usepackage{caption}
\usepackage{subcaption}
\usepackage{natbib}
\usepackage{soul}
\usepackage{lineno}

\begin{document}



\title{Superconducting integrated on-demand quantum memory with microwave pulse preservation}

\author{Aleksei R. Matanin}
\thanks{armatanin@bmstu.ru}
\author{Nikita S. Smirnov}
\affiliation{Shukhov Labs, Quantum Park, Bauman Moscow State Technical University, Moscow, 105005, Russia}
\affiliation{Dukhov Automatics Research Institute (VNIIA), Moscow 127030, Russia}

\author{Anton I. Ivanov}
\author{Victor I. Polozov}
\affiliation{Shukhov Labs, Quantum Park, Bauman Moscow State Technical University, Moscow, 105005, Russia}

\author{Daria A. Moskaleva}
\author{Elizaveta I. Malevannaya}
\affiliation{Shukhov Labs, Quantum Park, Bauman Moscow State Technical University, Moscow, 105005, Russia}
\author{Margarita V. Androschuk}
\author{Yulia A. Agafonova}
\author{Denis E. Shirokov}
\affiliation{Shukhov Labs, Quantum Park, Bauman Moscow State Technical University, Moscow, 105005, Russia}
\author{Aleksander V. Andriyash}
\affiliation{Dukhov Automatics Research Institute (VNIIA), Moscow 127030, Russia}

\author{Ilya A. Rodionov}
\thanks{irodionov@bmstu.ru}
\affiliation{Shukhov Labs, Quantum Park, Bauman Moscow State Technical University, Moscow, 105005, Russia}
\affiliation{Dukhov Automatics Research Institute (VNIIA), Moscow 127030, Russia}

\date{\today}

\begin{abstract}

Microwave quantum memory represents a critical component for quantum radars and resource-efficient approaches to quantum error correction.
Superconducting microwave resonators provide highly efficient storage, long coherence times, on-demand reading and even in memory pulse engineering, but it’s still challenging to overcome design and materials induced loss channels for on-chip realization. 
In this work, we present a novel architecture of integrated superconducting quantum memory with a dynamically controlled RF-SQUID coupling element in pulse regime, thus ensuring high efficiency storage and cycling storage time. 
It demonstrates a memory cycle time of 1.51 $\mu s$ and 57.5(4)\% storage fidelity with preservation of the stored pulse shape during the retrieval at single-photon level excitations. 
We establish that while the proposed active coupler realization introduces no measurable fidelity degradation, the primary limitation arises from impedance matching and materials imperfections. 
Still the device was used only for storing finite-duration near-single-photon classical microwave pulses, we assert that it operates as a linear device when the photon population in the common resonator remains low so it should be compatible with quantum state storage.
The proposed architecture highlights a disruptive potential for on-chip qubit and memory integration for scalable quantum error correction, while identifying specific avenues for near-unity storage fidelity.

\end{abstract}

\maketitle

Fault-tolerant quantum computing and quantum internet require quantum memory as an essential building block of a future quantum information processing platform \cite{kimble2008, Lvovsky2009, Wehner2018, Blais2021, Matteo2011}. Superconducting circuits quantum electrodynamics (cQED) is among the leading realizations of intermediate-scale quantum computers \cite{Morten2020, Devoret2013}.
Meanwhile there is a strong motivation to break the wall of nearest-neighbor qubit coupling using enhanced cQED architecture with integrated quantum memory \cite{Matteo2011, 2017-Naik-NatureComm, Pfaff2017, Axline2018,  Morten2020}. 
Moreover, it would allow to extend limited coherence time of the superconducting qubits, implement complex quantum algorithms \cite{QSearch-Lloyd-PRL-2008} and hardware-efficient quantum error correction \cite{Leghtas2013, Corcoles2015, Ofek2016, Rosenblum2018}. 
Compared with traditional superconducting qubits, high quality factor resonators have a superior potential for quantum state storage due to their impressive lifetime \cite{Ofek2016, Reagor2016, Wenner2014, Kobe2017}, efficient thermalization, no extra fridge control lines and ability to couple multiple qubits
\cite{2017-Naik-NatureComm, Kubo2011}.

The multi-resonator approach \cite{2018-Moiseev-SR} exploits the ideas of photon/spin echo \cite{Moiseev2001, Riedmatten2008,
Tittel2010,
Grezes2015,
Ranjan2020,
Moiseev_2021} 
in a system of resonators with a linear periodic spacing of their resonant frequencies. 
An echo forms in such resonators and reemits an input pulse after $\tau=1/\Delta$, where $\Delta $ is the frequency spacing between resonators.
In this case, the effective bandwidth of the memory is  determined by the span of the formed frequency comb, which significantly exceeds the linewidth of an individual resonator. 
Recently, a superconducting on-chip quantum memory device with fixed storage time based on four coplanar resonators was demonstrated with efficiency of about 73\% and 60\% in high-power and single-photon regimes, respectively \cite{matanin2023toward}.

There are several strategies for realizing quantum memory with on-demand retrieval in multi-resonator systems. The first approach relies on the dynamic adjustment of internal resonator frequencies \cite{2021-Bao-PRL}. 
In this method, the storage stage is implemented by detuning the resonator frequencies from the input pulse frequency.
Efficient information retrieval is enabled by re-tuning resonator frequencies back to the initial values.
This concept was experimentally validated using a system comprising four planar superconducting resonators, each integrated with a DC-SQUID for frequency control \cite{2021-Bao-PRL}. 
The device achieved a storage efficiency of approximately 12\%. 
The dominant contributions to infidelity originated from enhanced intrinsic resonator losses, predominantly due to the incorporation of DC-SQUIDs, along with the operation of the circuit in a transmission-mode configuration.

An alternative method for implementing on-demand quantum memory involves the creation of a virtual delay line by engineering the system's response to mimic that of a physical delay line through parametric control. This approach was experimentally demonstrated in a device comprising seven coplanar superconducting resonators coupled via an Asymmetrically Threaded SQUID (ATS) \cite{makihara2024}, enabling parametric modulation through a three-wave mixing process \cite{lescanne2020exponential}. 
The system achieved a storage fidelity of approximately 21\% for a storage duration of 2 $
\mu s$, with performance primarily limited by intrinsic losses in the ATS and the superconducting resonators.
However, this implementation requires three additional dedicated on-chip control lines, sensitive to impedance mismatches in the input-output transmission line and necessitates the simultaneous calibration of multiple parametric pumping parameters to achieve an ideal delay line profile. 
Moreover, strong parametric pump fields may introduce spurious noise, potentially compromising coherence.

\begin{figure*}
    \centering
    \includegraphics[width=0.95\linewidth]{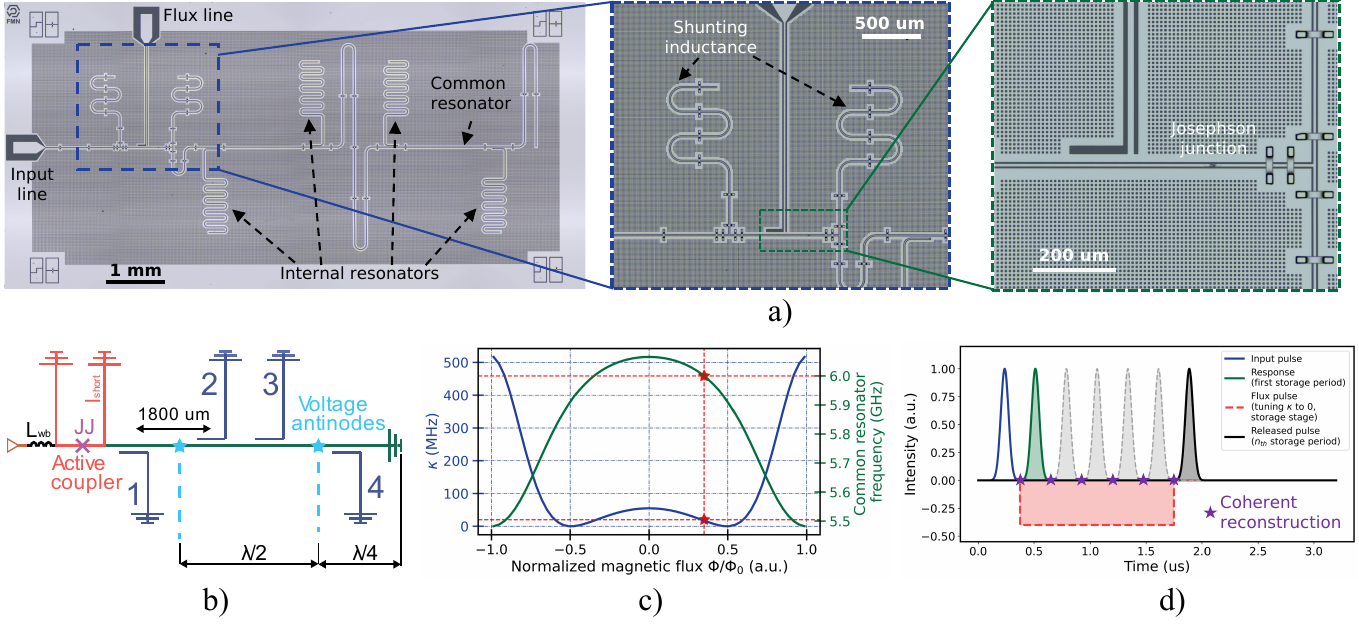}
    \caption{
    \textbf{(a)} Image of the fabricated quantum memory device.
    All structures are embedded in vortex-pinning hole array that helps achieve high internal Q-factor by magnetic vortices trapping \cite{McRae2020}
    \textbf{(b)} Principal scheme of the quantum memory device with the relative positions of the common resonator and internal resonators. The device is mounted in a sample
    holder and electrically connected to a copper printed
    circuit board (PCB) via three parallel aluminum wire
    bonds, introducing an additional parasitic inductance of
    $L_{wb}\approx 0.5\ nH$.
    \textbf{(c)} Modeled dependencies of $\kappa$ and common resonator frequency on external magnetic flux in RF-SQUID loop. Designed operating points with common resonator frequency of 6 GHz and $\kappa$ of 20 MHz (red dashed lines) correspond to magnetic flux of $\approx0.33$ and are marked by red stars.
    \textbf{(d)} Concept of storage stage implementation. Green pulse - device response without coupling strength tuning. At the moment of the dark state in the common resonator, the coupling between the external waveguide and the common resonator should be switched off for the required period of time (red shaded rectangular). At the n-th period of dark state in common resonator (purple stars) coupling strength could be tuned to impedance-matching value to effectively release quantum pulse from memory cell.
    }
    
    \label{fig:fig1}
\end{figure*}

Here, we present a novel architecture for on-demand quantum memory utilizing an active coupler based on RF-SQUID with low-loss single Josephson junction (JJ) \cite{geller2015, moskalev2023optimization, pishchimova2023improving}.
This design experimentally demonstrates periodic coherent storage with preservation of the input pulse shape in the near-single-photon regime, achieving a storage fidelity of 57.5(4)\% in the first retrieval cycle. 
The system exhibits an effective fidelity decay time of 11.44(3.58) $\mu s$.
The proposed system demonstrates a linear response regime under conditions of low photon population within the common resonator.
Despite that the device was used only for storing near-single-photon classical microwave pulses, we assert that it is compatible with quantum state storage. 
This claim is supported by prior demonstration of quantum properties in an analogous linear architecture without an active coupler \cite{matanin2023toward}.
Compared to previous realizations described above \cite{2021-Bao-PRL, makihara2024}, the proposed device offers several key advantages: it requires only one additional control line, fewer room-temperature electronics and the active coupler introduces no additional internal losses at the storage stage, thereby eliminating a major fidelity limitation observed in alternative architectures. Moreover, the active coupler exhibits straightforward integration compatibility with transmon qubits \cite{niu2023low, grebel2024bidirectional}.

Our quantum storage device (Fig. \ref{fig:fig1}\textbf{a}) consists of one common resonator that interacts with four internal high-Q superconducting resonators \cite{zikiy2023high} and coupling waveguide as it is depicted at principal scheme in Fig. \ref{fig:fig1}\textbf{b}.
The common resonator has a designed maximal frequency of $f_c = 6.06$ GHz with effective electrical length of $4\cdot\lambda/4$, where $\lambda$ is the resonant wavelength. 
Each of the four internal resonators has a length of $\sim\lambda/4$ and is coupled to the common resonator with coupling constant $g\approx 4.85$ MHz. 
The designed frequencies of the internal resonators range from 5.991 GHz to 6.009 GHz with a step of $\Delta=6$ MHz.
The minimal distance between the adjacent resonators is set to 1800 $\mu m$ (Fig. \ref{fig:fig1}\textbf{a}-\textbf{b}) to avoid parasitic coupling, which could otherwise induce an uncontrolled shift in their resonant frequencies and alter the effective loaded linewidth \cite{matanin2023toward}.
We also keep the position of each internal resonator at the same distance from the voltage antinodes of the common resonator, as depicted in Fig. \ref{fig:fig1}b. 
Such positioning allows providing an identical coupling constant for all internal resonators since coupling depends only on a mutual capacitance that is fabricated with good precision. 
The device is mounted in a sample holder and electrically connected to a copper printed circuit board (PCB) via three parallel aluminum wire bonds, introducing an additional parasitic inductance of $L_{wb}\approx0.5\ nH$.

An active coupling element is introduced between the common resonator and the external waveguide to enable tunability of the coupling constant 
$\kappa$. 
The active coupling element is implemented as a single Josephson junction shunted to ground by inductance at both ends (RF-SQUID) \cite{geller2015}.
The latter was chosen due to its compatibility with galvanic connections and superconducting transmon qubits \cite{geller2015}.
Moreover, this configuration enables complete decoupling of the quantum memory unit from the external waveguide, achieving 
$\kappa = 0$ MHz — a critical condition for high-efficiency storage, as will be discussed in subsequent sections.
Physically, RF-SQUID operates as a tunable inductive element under the condition that the AC current through the Josephson junction remains substantially below its critical current. 
The shunting inductors are made as coplanar waveguides with the same length at both ends of JJ, as it is depicted in Fig. \ref{fig:fig1}\textbf{a},\textbf{b}. 
The coupling strength $\kappa$ and common resonator frequency are adjusted simultaneously by varying the magnetic flux in RF-SQUID loop, see modeling results in Fig. \ref{fig:fig1}\textbf{c}. 

The quantum memory operation principle consist of the following three stages: (1) sending an input pulse to the quantum memory, (2) quantum state storage, and (3) controlled information retrieval from the memory cell to the external waveguide.
For the efficient transfer of an input pulse into the quantum memory cell it is crucial to ensure the impedance matching condition. 
For resonator's internal loss rates $\gamma\ll g,\kappa$ this condition relates $g$, $\Delta$  and $\kappa$ in the following expression (\ref{eq::IMC}):

\begin{flalign}
    \label{eq::IMC}
    \kappa=2\pi\frac{g^{2}}{\Delta}
\end{flalign}

According to the equation above, for the chosen values of $g=4.38$ MHz and $\Delta=6$ MHz, the value of $\kappa \approx20$ MHz at a common resonator frequency of 6 GHz.
Both of the $\kappa$ and $\Delta$ are determined by JJ's inductance, length of shorted sections $l_{short}$, and external magnetic flux (see Fig. \ref{fig:fig1}\textbf{d,e}).

The storage stage is implemented by maintaining zero coupling strength ($\kappa=0$) when the field amplitude in the common resonator is approximately zero (during dark state in the common resonator).
Figure \ref{fig:fig1}\textbf{c} demonstrates that tuning the coupling strength $\kappa$ induces an approximately 100 MHz shift of the common resonator frequency compared to that at the recording and retrieval stages, effectively decoupling it from the internal resonators. resulting in its decoupling from the internal resonators. 
Consequently, during the storage stage, the input field becomes predominantly localized within the internal resonators. 
This isolation mechanism ensures that even substantial losses in the common resonator do not compromise the device fidelity, as it is discussed in supplementary materials \cite{seeSupp} \nocite{Gardiner1985}. 
Under ideal conditions of equally spaced internal resonator frequencies, coherent superposition of all resonator's fields exhibits temporal periodicity of $T_{stor}=1/\Delta \approx166$ ns for designed parameters.
The release stage is initiated at the n-th constructive interference period by adjusting $\kappa$ to the impedance-matched value (see Fig. \ref{fig:fig1}\textbf{d}), enabling controlled information retrieval.

\begin{figure*}
    \includegraphics[width=0.94\linewidth]{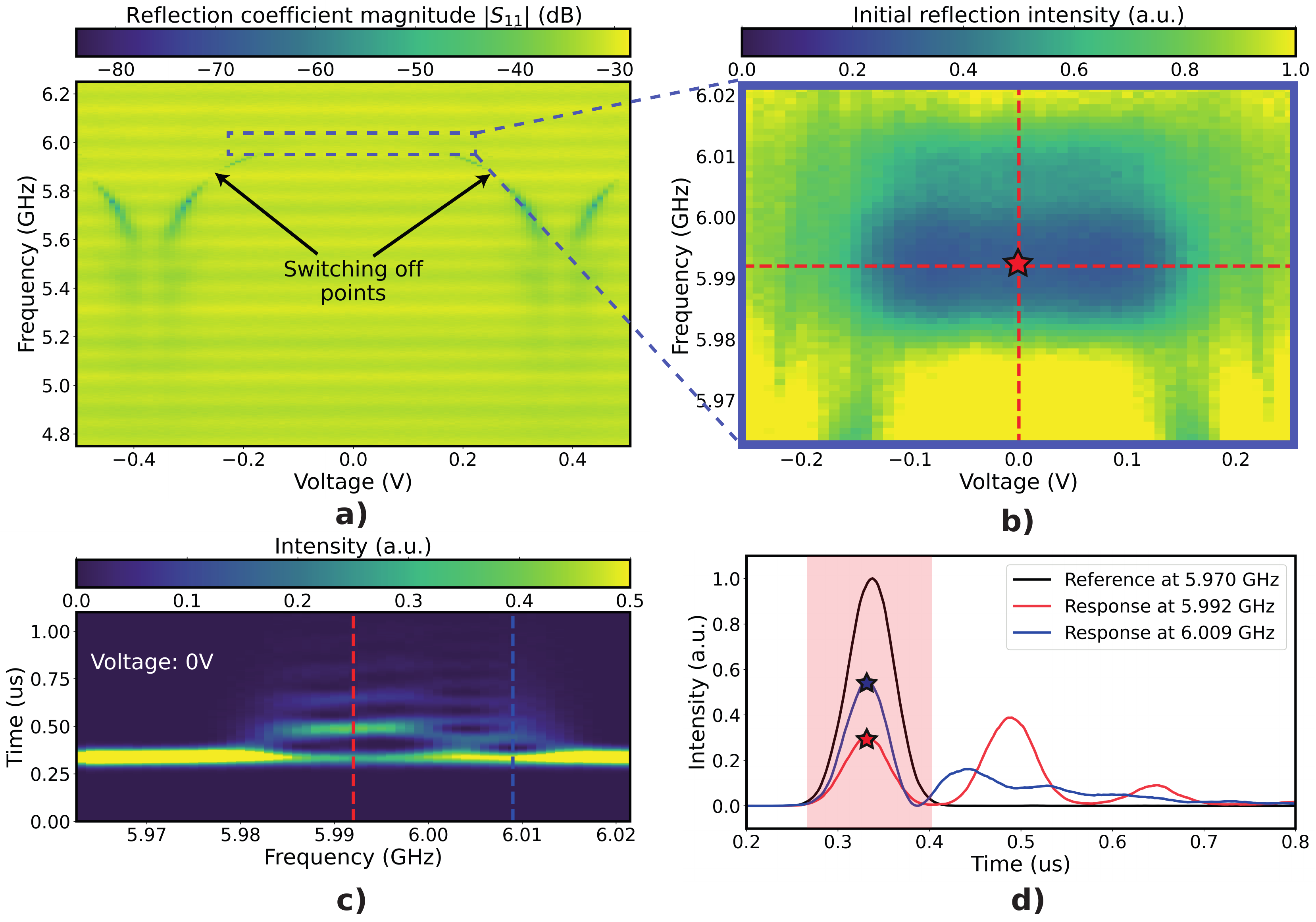}
    \caption{Device operation point calibration.
    \textbf{(a)} Device single-tone spectroscopy. Switching-off points marked by black arrows. 
    The voltage and pulse frequency fine range used in the next calibration stages is highlighted by a dashed blue frame.
    \cite{McRae2020}
    \textbf{(b)} initial reflection intensity dependence on voltage and pulse frequency in a fine range highlighted in \textbf{(a)}. A Frequency of 5.992 GHz and a voltage of 0V are chosen as operating points for record and release stages, such that they provide minimal initial reflection intensity.  
    \textbf{(c)} Impulse response intensity on pulse frequency at constant voltage of 0V. 
    \textbf{(d)} Reference (black) and two cross-sections marked by dashed lines at \textbf{(c)}.
    We define the initial reflection intensity as the maximal value (marked by corresponding stars) in the red shaded area.
    }
    \label{fig:fig6}
\end{figure*}

The experimental setup for characterizing the quantum memory device employs a configuration similar to that detailed in Ref. \cite{matanin2023toward}. 
Key components include a custom-built, infrared-blocking filter based on Eccosorb \cite{ivanov2023robust} and a quantum-limited impedance-matched Josephson parametric amplifier (IMPA) \cite{moskaleva2024lumped}. 
Following a down-conversion scheme, the in-phase, 
I(t), and quadrature, Q(t), signal components from the respective mixer channels are digitized using fast speed analog-to-digital converter for subsequent processing and analysis.
The coupling strength $\kappa$ is dynamically tuned by applying a voltage pulse from an arbitrary waveform generator to a dedicated flux line. This line includes low-pass and infrared (IR) filtering to suppress high-frequency noise and spurious thermal radiation and prevent standing waves \cite{Malevannaya2025Engineering, krinner2019engineering}.
Further experimental details are provided in the supplementary materials \cite{seeSupp}.

Fig. \ref{fig:fig6} presents the complete calibration results for the quantum memory device.
The device's single-tone spectroscopy (Fig. \ref{fig:fig6}\textbf{a}) reveals a maximum frequency shift of approximately 65 MHz from the designed resonator frequency, see supplementary materials \cite{seeSupp} for more details, resulting in compromised impedance-matching conditions.
A bias voltage of -0.22 V is identified as the switching-off point, where the device operates in complete reflection mode. 
This voltage value serves as the reference for subsequent pulse-frequency calibrations.
All subsequent calibration measurements are performed in pulsed operation mode using Gaussian-envelope pulses with full width at half maximum (FWHM) of 57 ns, which corresponds to spectral bandwidth of 15.5 MHz.
The pulse amplitude is chosen so that the resulting AC current is about one order less than the JJ's critical current, which in our case is tens of nanoamperes, to suppress higher-order nonlinear effects.
Under these conditions, the system operates in the near-single-photon regime, as confirmed by the negligible contribution from nonlinear processes. 
Indeed, the on-chip input power under pulsed operation is estimated as $P_{in}\approx-135\  dBm$, accounting for both room-temperature and cryogenic attenuation. 
For a Gaussian pulse with the FWHM duration of 57 ns, this corresponds to a mean photon number of $\braket{n}\approx4$.
Figure \ref{fig:fig6}\textbf{c} shows the normalized response intensity $\vert I(t)+1j\cdot Q(t)\vert^2$ as a function of pulse frequency at fixed bias voltage. 
The frequency range (from 5.965 GHz to 6.020 GHz, blue frame in Fig. \ref{fig:fig6}\textbf{a}) was selected based on preliminary single-tone spectroscopy. Normalization was performed such that $max(\vert I(t)+1j\cdot Q(t)\vert^2)$ = 1 for reference measurements, enabling quantitative efficiency analysis (note: this differs from the fidelity normalization protocol described subsequently).

In Fig. \ref{fig:fig6}\textbf{d} we demonstrate the normalized device response in the time domain for two distinct pulse frequencies (indicated by dashed lines in Fig. \ref{fig:fig6}\textbf{c}). 
The initial reflection intensity is defined as the maximum response within the reference time window (shaded red region, Fig. \ref{fig:fig6}\textbf{d}), with the corresponding values for both cases marked by stars.
Fig. \ref{fig:fig6}\textbf{b} illustrates the dependence of the initial reflection intensity on applied voltage offsets and pulse frequencies. 
The minimum reflection intensity occurs at 0 V and 5.992 GHz (denoted by a star in Fig. \ref{fig:fig6}\textbf{b}), signifying optimal impedance matching. 
This operating point was consequently selected for efficient device operation during recording and release stages.

\begin{figure}
    \centering
    \includegraphics[width=0.93\linewidth]{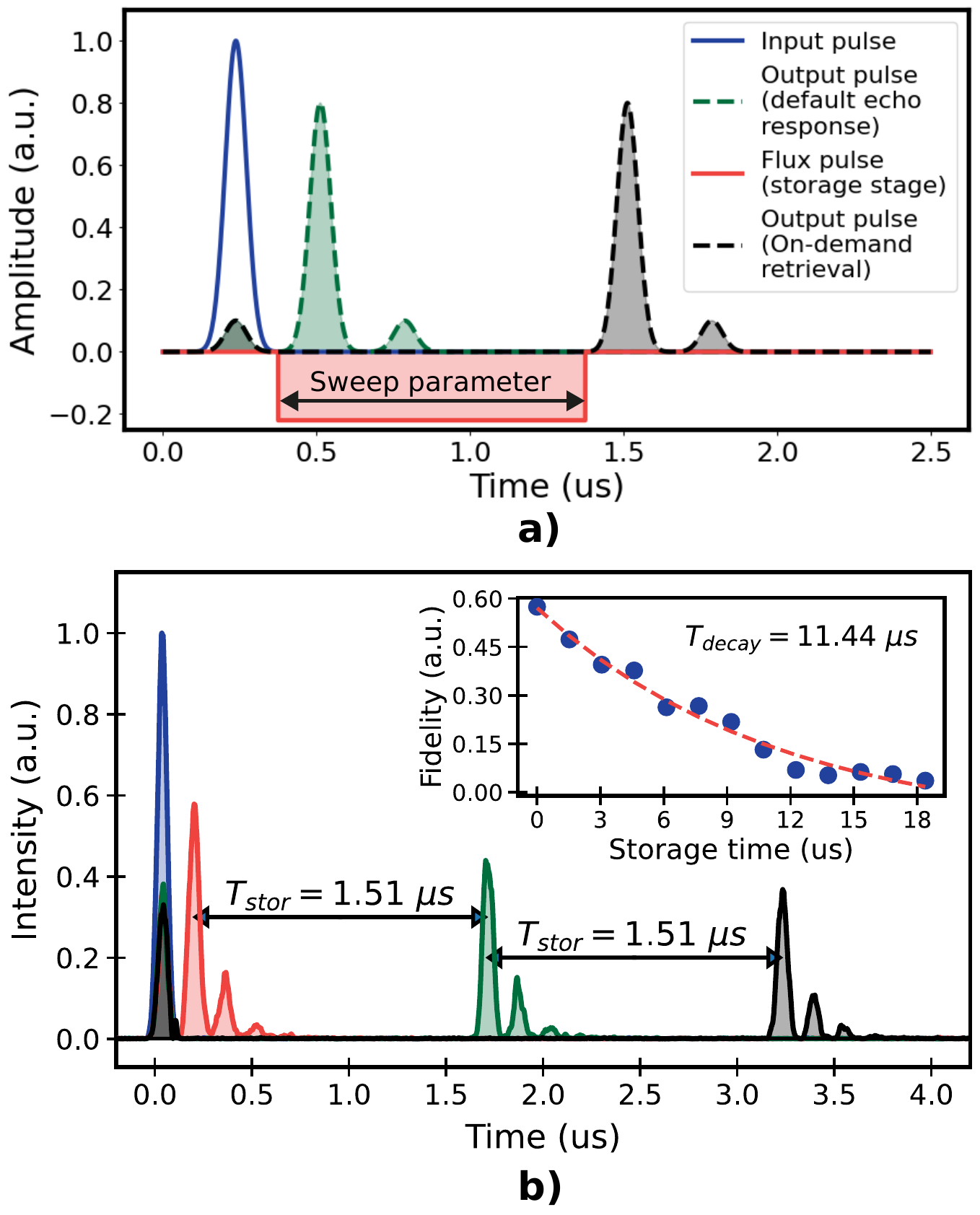}
    \caption{Device fidelity measurements.
    \textbf{a)} Pulse sequence for fidelity measurements. The storage stage duration is varied to demonstrate on-demand retrieval capability.
    \textbf{b)} Quantum memory responses for the first three storage periods (red, green and black, respectively. Blue graph corresponds to reference). We find memory cycle time equal to $T_{stor}=1.51\ \mu s$ and fidelity of first storage period $F=57.5(4)\%$.
    \textbf{Inset:} Fidelity on storage time. Due to internal losses,  it has exponential decay behavior with effective decay time $T_{decay}=11.44(3.58)\ \ \mu s$, which corresponds to effective quality factor of $Q_{i,eff}=4.3\cdot10^5$. 
    }
    \label{fig:fig3}
\end{figure}

We confirm the on-demand properties of the proposed scheme through pulsed microwave measurements, see Fig. \ref{fig:fig3}\textbf{a}.
Initially, a microwave pulse with a Gaussian envelope is input to the device. 
Subsequently, the active coupler is switched to its “closed” state by applying a magnetic flux pulse to the RF-SQUID. 
The duration of this flux pulse is systematically varied. 
The pulse amplitude required to achieve the off-state, determined from prior calibration, is -0.22 V.
In Fig. \ref{fig:fig3}\textbf{b} we present the normalized output intensity in the time domain for varying storage durations.
We observe that the retrieval fidelity with preservation of input pulse shape reaches a maximum cyclically, with a period of $T_{stor}=1.51\ \mu s$, which we define as the "cyclic time".
We assume that deviation of measured cycling time from the designed value of $T_{stor}=1/\Delta\approx166$ ns arises from non-equidistant spacing of the internal resonator frequencies. Quantitative analysis reveals a frequency spacing irregularity of $\delta = 660$ kHz (see Supplementary material \cite{seeSupp} for furher details), where the derived characteristic time $1/\delta=1.515\ \mu s$ shows excellent agreement with the observed storage period duration.
The memory device attains an efficiency of 57.9(2)\%, defined as the peak normalized intensity of the retrieved signal \cite{matanin2023toward}. 
Along with efficiency metric we employ a fidelity one \cite{makihara2024} to comprehensively characterize the quantum memory performance and ability of input pulse shape preservation (See supplementary material \cite{seeSupp} for further details). 
The measured fidelity of our device reaches 57.5(4)\% during the first release period, representing a several-fold improvement over previously reported values. 
The discrepancy between the measured efficiency and fidelity values is below 1\%, primarily attributed to distortions in the response pulse shape. 
These findings substantiate our assertion that the storage device effectively preserves the temporal profile of the input pulse.
Analysis reveals that normalized energy stored in the first storage period $\sum_{t}\,|g_{1}(t)|^2=0.991$, where $g_{1}(t)$ is the device’s temporal response mode, indicating the dominant contribution to the infidelity of impedance-matching imperfections.
The device exhibits a characteristic fidelity decay with storage time, as illustrated in inset of Fig. \ref{fig:fig3}, with an effective decay constant $T_{decay}=11.44\textcolor{red}{(3.58)} \ \mu s$. 
This corresponds to an effective internal resonator's quality factor in the "closed" regime of $Q_{i,eff}=4.3\cdot10^5$, consistent with independent measurements of the resonator’s internal quality factors (Qi) by fitting complex reflection coefficient $S_{11}$ \cite{probst2015efficient} measured with the tunable coupler in its "open" state (see supplementary information for details). 
A direct measurement of Qi with the coupler in the "closed" state is non-trivial. 
In this configuration, the resonators are deeply undercoupled (Qi $\ll$ Qc), which suppresses their resonant features in the global $S_{11}$ spectrum, precluding standard fitting methods. Based on this we claim that the internal quality factors of the internal resonators are not dependent on the operational state of the coupler.

In summary, we present a novel architecture for superconducting integrated on-demand quantum memory, which has been experimentally verified. 
This design offers significant advantages over existing implementations, requiring fewer control lines while eliminating the fidelity degradation typically caused by active coupling elements due to its decoupling from internal resonators during the storage stage. 
Our device demonstrates a memory cycle time of 1.51 $\mu s$ with 57.5(4)\% fidelity during the first release period and exhibits an effective fidelity decay time of 11.44(3.58) $\mu s$. 
The achieved storage fidelity represents a several-fold improvement over previous reports (21\% in Ref. \cite{makihara2024} and 12\% in Ref. \cite{2021-Bao-PRL}).
Detailed analysis identifies impedance-matching imperfections as the primary fidelity limitation, arising from non-equidistant resonator frequencies and deviations from the designed common resonator frequency. 
Future work will focus on addressing these specific challenges to further enhance device performance.

\textit{Acknowledgments -} We appreciate S.A. Moiseev, E.S. Moiseev, K.I. Gerasimov and N.S. Perminov from Kazan Quantum Center for fruitful discussions.
Device was fabricated at the BMSTU Nanofabrication Facility (Functional Micro/Nanosystems, FMNS REC, ID 74300).


\bibliography{QM_ActCoupl.bib}

\end{document}